\documentclass[12pt]{article}
\usepackage{amssymb,graphicx,color,epsfig,epic,pictex}
\begin{document}

\baselineskip=.22in
\renewcommand{\baselinestretch}{1.2}
\renewcommand{\theequation}{\thesection.\arabic{equation}}

\begin{flushright}
{\tt hep-th/0703144}
\end{flushright}

\vspace{5mm}

\begin{center}
{{\Large \bf BPS Limit of Multi- D- and DF-strings\\[2mm]
in Boundary String Field Theory
}\\[12mm]
Gyungchoon Go,~~Akira Ishida,~~Yoonbai Kim\\[2mm]
{\it Department of Physics, Sungkyunkwan University, Suwon 440-746, Korea}\\
{\tt gcgo, ishida, yoonbai@skku.edu}
}
\end{center}

\vspace{5mm}

\begin{abstract}
A BPS limit is systematically derived for straight multi- D- and
DF-strings from the D3$\bar{\mathrm{D}}3$ system in the context of
boundary superstring field theory. The BPS limit is obtained in the
limit of thin D(F)-strings, where the Bogomolnyi equation supports
singular static multi-D(F)-string solutions. For the BPS
multi-string configurations with arbitrary separations, BPS sum rule
is fulfilled under a Gaussian type tachyon potential and reproduces
exactly the descent relation. For the DF-strings ($(p,q)$-strings),
the distribution of fundamental string charge density coincides with
its energy density and the Hamiltonian density takes the BPS formula
of square-root form.
\end{abstract}


\newpage

\section{Introduction}

When the system of D-brane and ${\bar {\rm D}}$-brane decays,
lower-dimensional D-branes of codimension-two are produced as the
representative nonperturbative degrees~\cite{Sen:2004nf}. When the
D3${\bar {\rm D}}$3 is considered, the D-strings or DF-strings
($(p,q)$-strings) are particularly intriguing as cosmic string
candidates~\cite{Copeland:2003bj,Dvali:2003zj}. As has been done for
the cosmic strings from the Nielsen-Olesen vortices in Abelian-Higgs
model, the straight strings saturating the Bogomolnyi
bound~\cite{Bogomolny:1975de} enable us to study various dynamical
issues analytically~\cite{VS}.

The tachyon dynamics for D3${\bar {\rm D}}$3 is described in the
several contexts, and the boundary string field theory (BSFT or
background-independent string field theory)~\cite{Witten:1992qy}
should be an appropriate language with taking into account string
off-shell contributions~\cite{Gerasimov:2000zp}. In BSFT of D$p{\bar
{\rm D}}p$ for superstring theory, the effective BSFT action for a
complex tachyon field was derived and the descent relation for
single codimension-two brane was obtained in an exact form from the
energy density difference between the false and true
vacua~\cite{Kraus:2000nj}.

Since the kinetic term of the BSFT action is very complicated, the
static multi-D-string configurations and the related issues have
been dealt in the limited
references~\cite{Jones:2002si,Jones:2003ae,Jones:2003ew}. For BPS
static kinks and rolling tachyons in the BSFT of an unstable
D-brane, the equations of motion from the BSFT action were
analyzed~\cite{Hashimoto:2001rk,Sugimoto:2002fp,Kim:2006mg} and even
an exact topological BPS kink solution was
obtained~\cite{Kim:2006mg}. Additionally, in the Dirac-Born-Infeld
(DBI) type effective field theory (EFT) of a complex tachyon field
and U(1)$\times$U(1) gauge fields~\cite{Sen:2003tm,Garousi:2004rd},
some studies have been made recently. The single thin BPS vortex
satisfying the descent relation was reproduced~\cite{Sen:2003tm},
the solutions corresponding thick D- and DF-strings were found in
the presence of radial electric field~\cite{Kim:2005tw}, and the
gravitating solutions including black brane structure were
obtained~\cite{Kim:2006xi}. In relation with cosmic strings, the BPS
limit for static straight multi-D(F)-strings was
established~\cite{Kim:2006xi}.

In this Letter, we will consider the D$p{\bar {\rm D}}p$ action in
super-BSFT and derive rigorously a BPS limit for static straight
multi- D- and DF-strings. To be specific, the BPS limit is achieved
in the limit of zero thickness, the pressure components and
off-diagonal stress component vanish in the plane orthogonal to string
direction, a BPS sum rule based on the descent relation of
codimension-two branes is satisfied under a Gaussian-type tachyon
potential. The form of first-order Bogomolnyi equation is the same
as that in DBI-type EFT, and the multi-BPS-D(F)-string solutions
also satisfy the Euler-Lagrange equation. The obtained BPS
properties may open new windows to tackle dynamical and cosmological
issues with the
D(F)-strings~\cite{Jackson:2004zg,Tye:2005fn,Firouzjahi:2006vp} in
BSFT.

In section~\ref{sec2}, we derive the BPS limit for
multi-D(F)-strings in the context of BSFT. In section~\ref{sec3}, we
show that the Euler-Lagrange equation for the tachyon field does not
support static regular topological vortex solution, which may imply
uniqueness of singular BPS solutions as static D-vortex solutions.
We conclude in section~\ref{sec4} with brief discussions on further
studies.

\setcounter{equation}{0}
\section{BPS Multi- D- and DF-strings}\label{sec2}

In BSFT for superstrings, off-shell BSFT action $S$ is obtained through
an identification with worldsheet partition function $Z$,
$S=Z$~\cite{Marino:2001qc}. For the system of D$p{\bar {\rm D}}p$ in their
coincidence limit, the BSFT action of the tachyon field $T$ and its complex
conjugate ${\bar T}$, coupled to an Abelian gauge field $A_{\mu}$ with
$F_{\mu\nu}=\partial_{\mu}A_{\nu}-\partial_{\nu}A_{\mu}$,
is given by~\cite{Kraus:2000nj}
\begin{equation}\label{ac9}
S=-2{\cal T}_p \int d^{p+1}x \,V(T,{\bar T})
\sqrt{-\det (\eta_{\mu\nu}+F_{\mu\nu})}\,{\mathcal F}(y_+){\mathcal F}(y_-),
\end{equation}
where ${\mathcal T}_{p}$ is tension of the D$p$-brane. The runaway tachyon
potential is Gaussian type,
\begin{equation}\label{tpo}
V(T,{\bar T})=e^{-T{\bar T}},
\end{equation}
and functional form of the derivative term is
\begin{equation}\label{ktr}
{\mathcal F}(y_{\pm})=\frac{y_{\pm}\,4^{y_{\pm}}\,
\Gamma(y_{\pm})^2}{2\Gamma(2y_{\pm})},
\end{equation}
where the variables,
\begin{equation}\label{ypm}
y_{\pm}=2(G^{\mu\nu}\partial_\mu T \partial_\nu {\bar T})
\pm 2\sqrt{(G^{\mu\nu}\partial_\mu T \partial_\nu T)
(G^{\rho\sigma}\partial_\rho {\bar T} \partial_\sigma {\bar T})
+(\theta^{\mu\nu}\partial_\mu T \partial_\nu {\bar T})^2} \, ,
\end{equation}
are expressed in terms of
open string metric $G^{\mu\nu}$ and
noncommutativity parameter $\theta^{\mu\nu}$ as
\begin{equation}\label{gth}
G^{\mu\nu}=\left(\frac{1}{\eta+F}\right)^{(\mu\nu)},\qquad
\theta^{\mu\nu}=\left(\frac{1}{\eta+F}\right)^{[\mu\nu]} \, .
\end{equation}

Let us consider static multi-D(F)-strings from D3${\bar {\rm D}}$3
($p=3$), which are stretched parallel to $z$-direction. An
appropriate ansatz for the D-string and fundamental string is
\begin{equation}\label{ans}
T=T(x^{i}),\quad -F_{0z}=E_{z}(x^{i}), \quad (i=1,2),
\end{equation}
where all the other components of the field strength are assumed to
be vanishing. Substitution of the electric field $E_{z}$ (\ref{ans})
into the Bianchi identity,
$\partial_{\mu}F_{\nu\rho}+\partial_{\nu}F_{\rho\mu}
+\partial_{\rho}F_{\mu\nu}=0$, forces $E_{z}$ to be constant. The
static tachyon field (\ref{ans}) with constant electric field
$E_{z}$ leads to tachyon equation,
\begin{eqnarray}
\lefteqn{ 2\partial_i \left\{ V \sqrt{1-E_z^2}
\left[ \mathcal{F}'(y_+) {\mathcal F}(y_-) \left(
\eta^{ij} \partial_j T +\frac{
\eta^{ij}\partial_j {\bar T} (\eta^{kl} \partial_k T \partial_l T)}
{\sqrt{|\eta^{ij}\partial_i T \partial_j T|^2}}
\right) \right. \right. } \nonumber\\
&& \left. \left. +{\mathcal F}(y_+) \mathcal{F}'(y_-) \left(
\eta^{ij} \partial_j T -\frac{
\eta^{ij}\partial_j {\bar T} (\eta^{kl} \partial_k T \partial_l T)}
{\sqrt{|\eta^{ij}\partial_i T \partial_j T|^2}} \right) \right] \right\}
= \sqrt{1-E_z^2}\,{\mathcal F}(y_+)
{\mathcal F}(y_-) \frac{\partial V}{\partial {\bar T}},
\label{stq}
\end{eqnarray}
where $y_{\pm}$ in (\ref{ypm}) reduce to
\begin{equation}\label{spm}
y_{\pm}=2(\eta^{ij}\partial_i T \partial_j {\bar T})
\pm 2\sqrt{|\eta^{ij}\partial_i T \partial_j T|^2} \;.
\end{equation}
These field configurations automatically satisfy
the equation of the gauge field $A_{\mu}$, $\partial_\mu \Pi^{\mu\nu}=0$,
where $\Pi^{\mu \nu}\equiv \partial{\mathcal L}/\partial(\partial_\mu A_\nu)$.
Since the momentum densities,
$T^{0i}$ and $T^{0z}$,
and some off-diagonal stress components, $T^{iz}=T^{zi}$, are
vanishing under the ansatz (\ref{ans}), the conservation of
energy-momentum tensor becomes
\begin{equation}\label{cos}
\partial_{j}T^{ji}=0,
\end{equation}
and it is equivalent to the tachyon equation (\ref{stq}) for nontrivial
configurations.

To investigate the BPS limit of the D(F)-strings, we examine the
pressure components perpendicular to the D(F)-strings
\begin{eqnarray}
T^{x}_{\; x}&=&-2{\cal T}_{3} V(T)\sqrt{1-E_z^2}\left\{
{\mathcal F}(y_+){\mathcal F}(y_-)+2[\mathcal{F}'(y_+)\mathcal{F}(y_-)
y_+^{xx}+\mathcal{F}(y_+)\mathcal{F}'(y_-) y_-^{xx}]
\right\} ,
\label{txx}\\
T^{y}_{\; y}&=&-2{\cal T}_{3} V(T)\sqrt{1-E_z^2}\left\{
{\mathcal F}(y_+){\mathcal F}(y_-)+2\left[\mathcal{F}'(y_+)\mathcal{F}(y_-)
y_+^{yy}+\mathcal{F}(y_+)\mathcal{F}'(y_-) y_-^{yy}\right]
\right\},
\label{tyy}
\end{eqnarray}
where $y_{\pm}^{ij}$ are defined by
\begin{equation}\label{yij}
y_{\pm}^{ij}=-2\partial_{(i}T \partial_{j)}\bar{T} \mp
\frac{(\partial_i T \partial_j T)(\partial_k \bar{T} \partial_k \bar{T})
+(\partial_i \bar{T} \partial_j \bar{T})(\partial_k T \partial_k T)}
{\sqrt{|\partial_k T \partial_k T|^2}}.
\end{equation}
As a necessary condition, pressure difference is required to vanish;
\begin{eqnarray}
T^{x}_{\; x}-T^{y}_{\; y}&=&8{\cal T}_{3}V(T)\sqrt{1-E_z^2}\,
(|\partial_x T|^2-|\partial_y T|^2) \Bigg\{
\left[\mathcal{F}'(y_+)\mathcal{F}(y_-)+\mathcal{F}(y_+)\mathcal{F}'(y_-)
\right]\nonumber\\
&&\qquad +\frac{\left[\mathcal{F}'(y_+)\mathcal{F}(y_-)
-\mathcal{F}(y_+)\mathcal{F}'(y_-)
\right](|\partial_x T|^2+|\partial_y T|^2)}
{\sqrt{|(\partial_x T)^2+(\partial_y T)^2|^2}}\Bigg\}\nonumber\\
&=&4{\cal T}_{3}V(T)\sqrt{1-E_z^2}\,
\left[(\overline{\partial_xT+i\partial_yT})(\partial_xT-i\partial_yT)
+(\overline{\partial_xT-i\partial_yT})(\partial_xT+i\partial_yT)\right]
\nonumber\\
&&\times\Bigg\{
\left[\mathcal{F}'(y_+)\mathcal{F}(y_-)+\mathcal{F}(y_+)\mathcal{F}'(y_-)
\right]\nonumber\\
&&\hspace{7mm}+\frac{\left[\mathcal{F}'(y_+)\mathcal{F}(y_-)
-\mathcal{F}(y_+)\mathcal{F}'(y_-) \right](|\partial_x
T|^2+|\partial_y T|^2)} {2\sqrt{|(\partial_x T)^2+(\partial_y T)^2|^2}}\Bigg\}
\label{x-y}\\
&=&0.
\end{eqnarray}

We read first-order Cauchy-Riemann equation as Bogomolnyi equation
from vanishing pressure difference (\ref{x-y})
\begin{equation}\label{Beq}
(\partial_x\pm i\partial_y)T=0,\qquad (\partial_x
\ln\tau=\pm\partial_y\chi~\textrm{and}~
\partial_y \ln\tau=\mp\partial_x\chi),
\end{equation}
where $T=\tau e^{i\chi}$.\footnote{We call the first-order
Cauchy-Riemann equation the Bogomolnyi equation since every BPS
D(F)-string configuration is a solution of this equation and the gauged version
of this equation was one of the Bogomolnyi equations in
$(2+1)$-dimensional Abelian-Higgs model and its analogues.}
By using (\ref{Beq}), we easily check that
the remaining off-diagonal stress component becomes automatically
zero;
\begin{eqnarray}
T^{x}_{\; y}&=&-4{\cal T}_{3} V(T)\sqrt{1-E_z^2}\,[
\mathcal{F}'(y_+)\mathcal{F}(y_-)
y_+^{xy}+\mathcal{F}(y_+)\mathcal{F}'(y_-) y_-^{xy}]
\label{xy0}\\
&=&4{\cal T}_{3} V(T)\sqrt{1-E_z^2}\,
(\partial_xT\partial_y\bar{T}+\partial_yT\partial_x\bar{T})\nonumber\\
&&\hspace{-2mm}\times\left\{[\mathcal{F}'(y_+)\mathcal{F}(y_-)
+\mathcal{F}(y_+)\mathcal{F}'(y_-)]
+\frac{[\mathcal{F}'(y_+)\mathcal{F}(y_-)
-\mathcal{F}(y_+)\mathcal{F}'(y_-)](|\partial_xT|^2+|\partial_yT|^2)}
{\sqrt{|(\partial_x T)^2+(\partial_y T)^2|^2}}
\right\}\nonumber\\
&=&2{\cal T}_{3} V(T)\sqrt{1-E_z^2}\,\left[
(\partial_xT\pm i\partial_yT)(\overline{\partial_yT\pm i\partial_xT})
+(\overline{\partial_yT\mp i\partial_xT})
(\partial_xT\mp i\partial_yT)\right]\nonumber\\
&&\hspace{-2mm}\times\left\{[\mathcal{F}'(y_+)\mathcal{F}(y_-)
+\mathcal{F}(y_+)\mathcal{F}'(y_-)]
+\frac{[\mathcal{F}'(y_+)\mathcal{F}(y_-)
-\mathcal{F}(y_+)\mathcal{F}'(y_-)](|\partial_xT|^2+|\partial_yT|^2)}
{\sqrt{|(\partial_x T)^2+(\partial_y T)^2|^2}}\right\}\nonumber\\
&\stackrel{(\ref{Beq})}{=}&0.
\label{txy0}
\end{eqnarray}

For the $n$ straight strings (anti-strings) spread arbitrarily
on the $(x,y)$-plane, the ansatz on the tachyon phase $\chi$ is
\begin{equation}\label{Bph}
\chi=\pm\sum_{p=1}^n\theta_p=\pm\sum_{p=1}^n \tan^{-1}\frac{y-y_p}{x-x_p}.
\end{equation}
Then the tachyon amplitude $\tau$ is obtained as an exact solution
of the Bogomolnyi equation (\ref{Beq}),
\begin{equation}\label{Bam}
\tau=\prod_{p=1}^n \tau_{\rm BPS} |{\bf x}-{\bf x}_p|.
\end{equation}
Inserting the BPS solutions (\ref{Bph})--(\ref{Bam}) into the formula
(\ref{spm}), we obtain
\begin{equation}
y\equiv y_{\pm}=4\partial_{x}{\bar T}\partial_{x}T=
4 \tau_{{\rm
BPS}}^{2}\prod_{p=1}^{n}(\tau_{{\rm BPS}}|{\bf x}-{\bf x}_p|)^2
\sum_{q,r=1}^n\frac{\cos \theta_{qr}} {(\tau_{{\rm BPS}}|{\bf
x}-{\bf x}_q|)\; (\tau_{{\rm BPS}}|{\bf x}-{\bf x}_r|)},\label{y}
\end{equation}
where $\theta_{qr}$ is the angle between two vectors, $({\bf x}-{\bf
x}_q)$ and $({\bf x}-{\bf x}_r)$. Substituting
(\ref{Bph})--(\ref{y}) into the pressure components
(\ref{txx})--(\ref{tyy}), we have $-T^{x}_{\; x}=-T^{y}_{\;
y}=2{\cal T}_{3}V\sqrt{1-E_{z}^{2}} \,{\mathcal F}({\mathcal
F}-2y{\mathcal F}')$. Therefore, the pressure components
(\ref{txx})--(\ref{tyy}) vanish only in the limit of zero thickness
of each vortex, $\tau_{{\rm BPS}}\rightarrow \infty$, due to the
rapidly-decaying tachyon potential $V(\tau)$ (\ref{tpo}) except for
the site of each vortex ${\bf x}={\bf x}_{p}$, i.e.,
$\displaystyle{\lim_{\tau_{{\rm BPS}\rightarrow\infty}}( -T^{x}_{\;
x})|_{{\bf x}={\bf x}_{p}} =\lim_{\tau_{{\rm
BPS}\rightarrow\infty}}( -T^{y}_{\; y})|_{{\bf x}={\bf x}_{p}}
=\frac{\pi}{2}{\cal T}_{3}\sqrt{1-E_{z}^{2}}}\,$. This nonvanishing
pressure at each D(F)-string location is different from the
character of BPS vortices in Abelian gauge theories with Higgs
mechanism where the pressure components vanish everywhere including
vortex points~\cite{Bogomolny:1975de,Hong:1990yh}. The stress
component $T^{x}_{\; y}$ also vanishes for the BPS configuration as
shown in (\ref{xy0})--(\ref{txy0}), and then the conservation of
energy-momentum tensor (\ref{cos}) reduces to $\partial_{x}T^{x}_{\;
x}=0$ and $\partial_{y}T^{y}_{\; y}=0$. For the aforementioned
pressure components of the BPS D(F)-strings in the infinite
$\tau_{{\rm BPS}}$ limit, the equations hold when the derivatives
are considered as weak derivatives~\cite{Eva}. As $\tau_{{\rm
BPS}}\rightarrow \infty$, the static singular solution
(\ref{Bph})--(\ref{Bam}) of BPS equation satisfies the conservation
of energy-momentum tensor (\ref{cos}), which is equivalent to the
tachyon equation (\ref{stq}) for nontrivial tachyon configurations.
In the section~\ref{sec3}, we also
show that the tachyon equation (\ref{stq}) does not support regular
static straight D(F)-string solution.

For the static configurations of $\dot{T}=\dot{\bar{T}}=0$ with constant
$E_{z}$, the conjugate momenta of the tachyon field and its complex conjugate
vanish, $\Pi_{T}\equiv \partial {\mathcal L}/\partial {\dot T}=0$
and $\Pi_{{\bar T}}\equiv \partial {\mathcal L}/\partial {\dot {\bar T}}=0$,
and the conjugate momentum of the gauge field $\Pi_{z}$ is
\begin{equation}\label{piz}
\Pi_{z}=-\frac{E_z}{\sqrt{1-E_z^2}} \;2{\cal T}_{3}V
{\mathcal F}(y_+){\mathcal F}(y_-).
\end{equation}
The Hamiltonian density obtained by a Legendre transform leads to
the BPS formula for DF-strings
(($p,q$)-strings)~\cite{Witten:1995im},
\begin{equation}\label{bpsH}
{\mathcal H}=\sqrt{\Pi_{z}^{2}+\left[2{\cal T}_{3}V
{\mathcal F}(y_{+}){\mathcal F}(y_{-})\right]^{2}}\, ,
\end{equation}
where the limit of D-strings,
${\mathcal H}|_{\Pi_{z}=0}=2{\cal T}_{3}
V{\mathcal F}(y_{+}){\mathcal F}(y_{-})$,
is trivially involved in the absence of fundamental
string charge density $\Pi_{z} = 0$.
Plugging the conjugate momentum (\ref{piz}), the Hamiltonian density
(\ref{bpsH}) coincides exactly with the energy density $-T^{t}_{\; t}$,
and, due to the boost symmetry along the $z$-direction,
the multi-D(F)-string configuration satisfies $T^{t}_{\; t}=T^{z}_{\; z}$;
\begin{eqnarray}\label{edt}
-T^{t}_{\; t}=-T^{z}_{\; z}=\frac{2{\cal T}_{3} V(T)}{\sqrt{1-E_z^2}}
{\cal F}(y_+){\cal F}(y_-).
\end{eqnarray}
Noticing easily that the energy density is proportional to the electric
flux density as
\begin{equation}
{\cal H}=-\frac{\Pi_{z}}{E_{z}},
\end{equation}
we read for the DF-strings that the charge
distribution of fundamental string part is exactly proportional to the
energy density of D-string part, which is confined at each string site in
$(x,y)$-plane.

If we require a BPS sum rule to the energy per unit D(F)-string length
for the BPS configuration with $y_{+}=y_{-}$,
\begin{eqnarray}\label{suru}
{\cal T}_{1}|n|=\frac{H}{\int_{-\infty}^{\infty} dz}
=\frac{2{\cal T}_{3}}{\sqrt{1-E_z^2}}\int d^{2}x\lim_{\tau_{{\rm BPS}}
\rightarrow \infty}V{\mathcal F}^{2}
=(2\pi\sqrt{\alpha'}\,)^{2}\frac{{\cal T}_{3}}{\sqrt{1-E_z^2}}|n|,
\quad (\alpha'=2),
\end{eqnarray}
the descent relation of a D(F)-string,
\begin{eqnarray}\label{dsc1}
{\cal T}_{1}=(2\pi\sqrt{\alpha'}\,)^{2}{\cal
T}_{3}/\sqrt{1-E_z^2}\,,
\end{eqnarray}
is correctly reproduced, and a constraint condition for a BPS sum rule
is achieved for the tachyon potential,
\begin{eqnarray}\label{Vco}
4\pi^{2}|n|=\int d^{2}x\lim_{\tau_{{\rm BPS}}\rightarrow \infty}V(\tau)
{\mathcal F}^{2}.
\end{eqnarray}
Since the integrand, $\displaystyle{\lim_{\tau_{{\rm
BPS}}\rightarrow \infty} V(\tau){\mathcal F}^{2}}$, has infinity at
each string site ${\bf x}={\bf x}_{p}$ and vanishes at ${\bf
x}\ne{\bf x}_{p}$ in the BPS limit of infinite $\tau_{{\rm BPS}}$,
the condition (\ref{Vco}) is reexpressed by a local form,
\begin{equation}\label{sde}
\lim_{\tau_{{\rm BPS}}\rightarrow\infty}
V(\tau){\mathcal F}^{2}=4\pi^{2}\sum_{p=1}^{n}\delta^{(2)}
({\bf x}-{\bf x}_{p}).
\end{equation}
In summary, the energy-momentum tensor of $n$ D(F)-strings is in the
BPS limit,
\begin{eqnarray}
-T^{\mu}_{\;\nu}={\cal T}_{1}\sum_{p=1}^{n}\,{\rm
diag}\,\left(\delta^{(2)} ({\bf x}-{\bf x}_{p}),
\frac{1-{E_{z}}^2}{16\pi}{\mathbb I} ({\bf x}-{\bf x}_{p}),
\frac{1-{E_{z}}^2}{16\pi}{\mathbb I} ({\bf x}-{\bf x}_{p}),
\delta^{(2)}({\bf x}-{\bf x}_{p})\right),
\end{eqnarray}
where ${\mathbb I} ({\bf x}-{\bf x}_{p})$ has unity at ${\bf x}={\bf x}_{p}$
and zero at ${\bf x}\ne {\bf x}_{p}$. 
Note that the pressure components
orthogonal to the string direction vanish in the limit of critical
electric field, $|E_{z}|\rightarrow 1$.

From now on, let us perform the integration (\ref{Vco}) with the
tachyon potential (\ref{tpo}) and show that it reproduces the
required value for saturating the BPS sum rule. First, we consider a
single D(F)-string of $n=1$ at an arbitrary position. In this case,
$y$ is independent of $x$, $y=4\tau_{\rm BPS}^2$, so is ${\mathcal
F}$. Then, a rescaling ${\tilde {\bf x}}=\tau_{{\rm BPS}}{\bf x}$
with a translation in (\ref{Vco}) provides a definite integral
without explicit dependence of $\tau_{{\rm BPS}}$;
\begin{eqnarray}\label{1i1}
\int d^2x \lim_{\tau_{{\rm BPS}}\rightarrow \infty}
e^{-(\tau_{\rm BPS}|{\bf x}|)^2} \mathcal{F}(4\tau_{\rm BPS}^2)^2
&=&\lim_{\tau_{{\rm BPS}}\rightarrow \infty}
\frac{\mathcal{F}(4\tau_{\rm BPS}^2)^2}{\tau_{\rm BPS}^2}
\int d^2{\tilde x}\, e^{-|{\tilde {\bf x}}|^2}.
\label{1i2}
\end{eqnarray}
If we perform the Gaussian integral for arbitrary $\tau_{{\rm BPS}}$
and take the limit of infinite $\tau_{{\rm BPS}}$ by using the
asymptotic form of $\mathcal{F}(y)^{2}$, $\mathcal{F}(y)^2=\pi
y+\pi/8+{\cal O}(y^{-1})$, in (\ref{1i1}), then value of the
integral is $4\pi^{2}$, which satisfies the descent relation.
Second, we consider the superimposed D(F)-strings of arbitrary
$|n|$. Now $y$ of ${\mathcal F}(y)$ has $x$-dependence as $y=4n^2
\tau_{\rm BPS}^2(\tau_{\rm BPS}|{\bf x}|)^{2n-2}$, and then we use
the same rescaling of ${\bf x}$ as
\begin{eqnarray}\label{aaa}
\lefteqn{ \int d^2x\lim_{\tau_{{\rm BPS}}\rightarrow \infty}
e^{-(\tau_{\rm BPS}|{\bf x}|)^{2n}}
\mathcal{F}(4n^2 \tau_{\rm BPS}^2(\tau_{\rm BPS}|{\bf x}|)^{2n-2})^2}\\
&=&4\pi n^2 \lim_{\tau_{{\rm BPS}}\rightarrow \infty}
\int d^2{\tilde x}\,e^{-|{\tilde {\bf x}}|^{2n}}|{\tilde {\bf x}}|^{2n-2}
\frac{\mathcal{F}(4n^2 \tau_{\rm BPS}^2|{\tilde {\bf x}}|^{2n-2})^2}{
4n^{2}\pi\tau_{\rm BPS}^{2}|{\tilde {\bf x}}|^{2n-2}}.
\label{ni1}
\end{eqnarray}
As $\tau_{{\rm BPS}}$ increases, the integrand with explicit
$\tau_{{\rm BPS}}$ dependence becomes
\begin{eqnarray}
\lefteqn{
\lim_{\tau_{{\rm BPS}}\rightarrow \infty}
\frac{\mathcal{F}(4n^2 \tau_{\rm BPS}^2|{\tilde {\bf x}}|^{2n-2})^2}{
4n^{2}\pi\tau_{\rm BPS}^{2}|{\tilde {\bf x}}|^{2n-2}}}
\nonumber\\
&=&\lim_{\tau_{{\rm BPS}}\rightarrow \infty}\left[
\frac{(\sqrt{\pi\times 4n^2 \tau_{\rm BPS}^2|{\tilde {\bf x}}|^{2n-2}}
\,)^2}{4n^{2}\pi\tau_{\rm BPS}^2|{\tilde {\bf x}}|^{2n-2}}
+{\mathcal O}(1/\tau_{{\rm BPS}}^{2},1/|{\tilde {\bf x}}|^{2n-2})\right]
\label{Fap}\\
&=&1
\label{ni9}
\end{eqnarray}
with keeping $|{\tilde {\bf x}}|$ finite. For infinite $|{\tilde
{\bf x}}|$, the integrand vanishes due to the exponential term.
Since ${\mathcal F}(y)^{2}$ is analytic for every non-negative $y$,
the integrand is finite at ${\tilde {\bf x}}=0$, and the asymptotic
form of $\mathcal{F}(y)^{2}$ guarantees finiteness of the integral
(\ref{ni1}) for finite $\tau_{{\rm BPS}}$, we can take infinite
$\tau_{{\rm BPS}}$ limit to ${\mathcal F}(y)^{2}/\pi y$ part in
(\ref{ni1}). Therefore, value of the integral (\ref{ni1}) is
$4\pi^{2}n$ which fits (\ref{Vco}). Third, we consider the case of
$n$ separated D(F)-strings where the distance between any pair of
D(F)-strings is much larger than $1/\tau_{{\rm BPS}}$. When ${\bf
x}\ne{\bf x}_s$ $(s=1,2,\ldots,n)$, it is obvious that $y$ in
(\ref{y}) diverges in the $\tau_{\rm BPS}\to\infty$ limit for any
tachyon field. When ${\bf x}={\bf x}_s$, the term with $p=q=r$ in
(\ref{y}) survives and hence $y\to\infty$ in this BPS limit. Thus we
see that $y$ always becomes infinite in the $\tau_{\rm
BPS}\to\infty$ limit. Accordingly, ${{\cal{F}}(y)}^2$ in the
integral (\ref{Vco}) diverges everywhere. Let us examine the tachyon
potential part in (\ref{Vco}). When ${\bf x}={\bf x}_p$
$(p=1,2,\ldots,n)$, $\tau$ in (\ref{Bam}) vanishes and then the
tachyon potential has unity, $V(\tau=0)=1$. When ${\bf x}\ne{\bf
x}_s$, it vanishes in the infinite $\tau_{\rm BPS}$ limit and the
integrand in (\ref{Vco}) also vanishes due to the exponential
damping of the tachyon potential despite of the leading divergent
term of $\mathcal{F}$, ${\mathcal F}(y)\rightarrow \sqrt{\pi y}\, $.
Therefore, among $n^2$-terms in $y$ (\ref{y}) specified by the $(q,
r)$-indices, the $n$-terms with $q=r$ contribute to the integral
(\ref{Vco}). In addition, functional shape of the integrand diverges
at each string site but vanishes away from the location of each
D(F)-string. In what follows, we will show that the contribution of
each term at ${\bf x}={\bf x}_p$ to the integration is exactly the
same as that of delta function given in single D(F)-string
(\ref{1i2}) as far as the distance $|{\bf x}_p-{\bf x}_q|$ for any
$p$ and $q$ ($p\ne q$) is sufficiently larger than $1/\tau_{{\rm
BPS}}$. Since only the neighborhoods of D(F)-string sites, ${\bf
x}={\bf x}_p$, contribute to (\ref{Vco}) in performing the
$(x,y)$-integration and become sufficiently small for infinite
$\tau_{\rm BPS}$, only the leading terms of $V$ and ${\cal{F}}^2$
can contribute nonvanishing value to the integral (\ref{Vco}). To be
specific, we can replace the integrand $V{\cal{F}}^2$ and then
perform the integration as follows,
\begin{eqnarray}
\lim_{\tau_{\rm BPS}\to\infty}\int d^2x V\mathcal{F}^2&=& 4\pi\int
d^2x \lim_{\tau_{\rm BPS}\to\infty} \sum_{s=1}^n
\exp\left[-\left(\prod\limits_{\scriptstyle p=1\atop\scriptstyle
(p\ne s)}^n \tau_{\rm BPS} |{\bf x}_s-{\bf x}_p|\right)^2
\left(\tau_{\rm BPS} |{\bf x}-{\bf x}_s|\right)^2
\right]\nonumber\\
&&\times\tau_{\rm BPS}^2 \left(\prod\limits_{\scriptstyle
q=1\atop\scriptstyle (q\ne s)}^n \tau_{\rm BPS} |{\bf x}_s-{\bf
x}_q|
\right)^2\nonumber\\
&=&4\pi^2n,
\end{eqnarray}
which is exactly the value in (\ref{Vco}). Fourth, we consider the
case of arbitrary BPS configuration where $n_{p}$ D(F)-strings among
the $n$ D(F)-strings are superimposed at an ${\bf x}_{p}$ with
$\displaystyle{n=\sum_{p}n_{p}}$. If we replace the integration
(\ref{1i2}) by (\ref{aaa})--(\ref{ni9}), the integration reproduces
the value in (\ref{Vco}) by applying repeatedly the above third
argument. In synthesis, the aforementioned four arguments lead to a
conclusion that the Gaussian type tachyon potential (\ref{tpo})
fulfills the integration (\ref{Vco}) in the thin BPS limit.

\setcounter{equation}{0}
\section{Nonexistence of Nonsingular D- and DF-string
Solutions}\label{sec3}

In this section, we deal with the tachyon equation (\ref{stq}) and
discuss nonexistence of the monotonically-increasing nonsingular
D-vortex solution connecting the boundary conditions at the origin,
$\tau(|{\bf x}|=0)=0$, and infinity, $\tau(|{\bf x}|=\infty)=\infty$.
This perhaps supports uniqueness of the singular BPS
multi-D(F)-string solutions obtained in the previous section.

Suppose that we have $n$ superimposed straight D(F)-strings
stretched along the $z$-axis. Since the electric field $E_{z}$ is
actually canceled in both sides of the tachyon equation (\ref{stq}),
we have
\begin{eqnarray} \label{Teq2}
\frac{1}{r} \frac{d}{dr}\left[r e^{-\tau^2}\tau'
{\mathcal F}'(y_+){\mathcal F}(y_-) \right] = \tau e^{-\tau^2}
{\mathcal F}(y_+)\left[ \frac{n^2}{r^2}{\mathcal F}'(y_-)
-\frac{1}{4}{\mathcal F}(y_-)\right] \,,
\end{eqnarray}
where $y_\pm$ in (\ref{ypm}) become
\begin{eqnarray}
y_+=4 \tau'^2\,,\qquad y_-=\frac{4n^2}{r^2}\tau^2 \,.
\end{eqnarray}
The D(F)-string solutions of our interest are given by
monotonically-increasing tachyon
configurations connecting the boundary conditions, $\tau(r=0)=0$ and
$\tau(r=\infty)=\infty$.

Expansion of the tachyon amplitude $\tau$ near the origin is
\begin{eqnarray}\label{rze}
\tau (r) \approx \tau_0 r^n (1-\tau_{1}r^{2}+...\, ),
\end{eqnarray}
where $\tau_0$ is an undetermined constant determined by the behavior at
asymptotic region. Since the coefficient of subleading term $\tau_{1}$ is
always positive irrespective $n$,
\begin{equation}
\tau_{1}=\left\{
\begin{array}{lc}
\displaystyle{\frac{{\mathcal F}(4\tau_{0}^{2})
-8\tau_{0}^{2}{\mathcal F}'(4\tau_{0}^{2})}{
32[{\mathcal F}'(4\tau_{0}^{2})+8\tau_{0}^{2}{\mathcal F}''(4\tau_{0}^{2})]}}
\, ,
& (n=1) \\
\displaystyle{ \frac{1}{24\ln 2}\left\{64\tau_{0}^{2}\left[8(\ln
2)^{2}-\frac{\pi^{2}}{3} \right]+\frac{1}{4}\right\}}\, ,
&(n=2)\\
\displaystyle{\frac{1}{32(\ln 2)(n+1)}}\, ,
&(n\ge 3)
\end{array}
\right. ,
\end{equation}
increasing tendency of the tachyon field $\tau(r)$ decreases as $r$ increases.
If we try expansion at asymptotic region by using a power law,
$\tau\sim \tau_\infty r^k,~(k>0)$, or a logarithmic increase,
$\tau\sim \tau_\infty \ln r$, many possibilities are ruled out by the
tachyon equation (\ref{Teq2}) and survived cases are
\begin{equation}\label{rinf}
\tau(r)\approx \tau_{\infty 0}r^{1+k}+\tau_{\infty 1}r^{1+k-l}+...,
\qquad (k>0,~0<l<2k),
\end{equation}
where both $\tau_{\infty 0}$ and $\tau_{\infty 1}$ are not
determined by the tachyon equation (\ref{Teq2}). The leading term is
rapidly increasing since $\displaystyle{\lim_{r\rightarrow
\infty}\tau'\rightarrow \infty}$.
 Comparison of the power
series solutions near the origin (\ref{rze}) and at the asymptotic
region (\ref{rinf}) suggests that smooth connection of both
increasing tachyon profiles seems unlikely.

\begin{figure}[h]
\begin{center}
\scalebox{1.3}[1.3]{\includegraphics{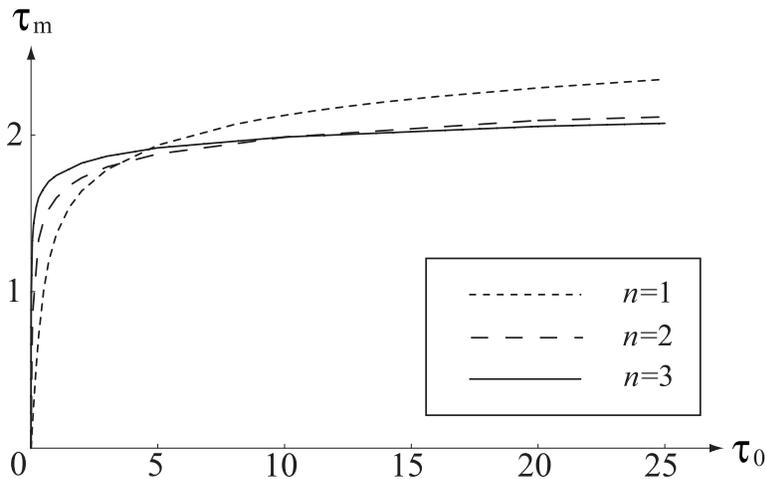}}
\par
\vskip-2.0cm{}
\end{center}
\caption{\small $\tau_{0}$ vs. $\tau_{{\rm m}}$. $n=1$ for dotted
line, $n=2$ for dashed line, and $n=3$ for solid line.}
\label{fig1}
\end{figure}

Another possibility is the solution with maximum value, i.e., the
tachyon amplitude increases near the origin, reaches a maximum value
$\tau_{{\rm m}}$ at a finite coordinate $r=r_{{\rm m}}$, and then
starts to decrease with $d^{2}\tau/dr^{2}|_{r=r_{{\rm m}}}<0$.
Expansion near $r=r_{{\rm m}}$ gives
\begin{equation}\label{rme}
\tau(r)\approx \tau_{{\rm m}}\left[1-\frac{1}{2}\tau_{{\rm
m}2}(r-r_{{\rm m}})^{2} +...\, \right],
\end{equation}
where the coefficient $\tau_{{\rm m}2}$ is
\begin{eqnarray}
\tau_{{\rm m}2}=\frac{{\mathcal F}(y_{-}^{{\rm m}})-\frac{4n^{2}}{
r_{{\rm m}}^{2}} {\mathcal F}'(y_{-}^{{\rm m}})}{8(\ln 2)
{\mathcal F}(y_{-}^{{\rm m}})},\qquad y_{-}^{{\rm
m}}=\frac{4n^{2}\tau_{{\rm m}}^{2}}{r_{{\rm m}}^{2}}.
\end{eqnarray}
In order to have the maximum $\tau_{\rm m}$=$\tau(r_{\rm m})$,
$\tau_{\rm m}$ and $r_{\rm m}$ should satisfy the following
inequality,
\begin{equation}\label{maxcon}
\tau_{\rm m}^2 > {\frac{d}{d({\rm ln}{y_-^{\rm m}})}\rm
ln{\mathcal F}}(y_-^{\rm m}).
\end{equation}
Numerical works support that every regular solution with finite
$\tau_{0}$ has the maximum value $\tau_{{\rm m}}$ at a finite
$r_{{\rm m}}$ irrespective of $n$ as shown in Fig.~\ref{fig1}.
Probably, there does not exist any static nonsingular
monotonically-increasing D(F)-string solution of the tachyon
equation (\ref{Teq2}) with $\tau(0)=0$ and $\tau(\infty)=\infty$.
Since the aforementioned discussion does not rule out the singular
solution $\tau$ with infinite slope $d\tau/dr\sim\infty$, the BPS
solutions (\ref{Bph})--(\ref{Bam}) are free from the argument of
nonexistence. This conclusion, the nonexistence of regular static
topological non-BPS D(F)-string solutions, is consistent with the
same result of nonexistence in DBI EFT~\cite{Kim:2005tw}.

\setcounter{equation}{0}
\section{Conclusion}\label{sec4}

The system of D3${\bar {\rm D}}$3 has been considered in the scheme
of super-BSFT EFT (\ref{ac9}) including a complex tachyon and
U(1)$\times$U(1) gauge fields. From the vanishing pressure
difference, the first-order Bogomolnyi equation (\ref{Beq}) was
derived and straight topological BPS multi-D(F)-string
configurations were given as exact static solutions
(\ref{Bph})--(\ref{Bam}) which also satisfy the conservation of
energy-momentum tensor (\ref{cos}). Since the forms of derived
Bogomolnyi equation and singular BPS solutions coincide exactly with
those in DBI type EFT, this BPS structure seems universal and is
consistent with type II superstring theories. The expression of
energy was rewritten by the BPS sum rule for the BPS
multi-D(F)-string solutions (\ref{suru}), and reproduced the descent
relation for codimension-two objects (\ref{dsc1}), which allowed to
interpret the obtained vortex-strings as BPS D1-branes in IIB string
theory. This results in a constraint condition for the BPS tachyon
potential (\ref{Vco}), and the Gaussian type potential of BSFT
(\ref{tpo}) fulfills the condition. Since it is nothing but making a
sum of delta functions in the thin BPS limit (\ref{sde}), the
uniqueness of BPS tachyon potential seems unlikely. When the
$z$-component of constant electric field (\ref{ans}) is turned on,
the conjugate momentum of the gauge field, the charge density of
fundamental strings (\ref{piz}), is confined along the D-strings. In
addition, the corresponding Hamiltonian density takes a BPS formula
(\ref{bpsH}), $\sqrt{p^{2}+q^{2}}$ form for the D1-charge density
$q$ and the fundamental string charge density $p$, so that the
configuration with constant electric field along the string
direction is the DF-string (or $(p,q)$-string) from
D3$\bar{\mathrm{D}}3$. Though we checked the conditions for BPS
vortex configurations explicitly, the form of obtained BPS limit
is different from the usual BPS bound for vortices, of which energy
minimum is saturated only when the Bogomolnyi equations are satisfied.
In this sense, the BPS bound for codimension-two branes from D${\bar {\rm D}}$
system needs further
study. We also checked the possibility that the
tachyon equation (\ref{Teq2}) could possess a nonsingular
D(F)-string solutions and the analysis supported negative answer.

Since we achieved a BPS limit of multi-vortex-strings, it may open
systematic study of classical dynamics of BPS multi-D(F)-strings,
particularly moduli space dynamics in the context of BSFT. Studies
of the D(F)-strings in curved spacetime naturally have cosmological
implication as candidates of cosmic superstrings.

\section*{Acknowledgments}
We would like to thank Dongho Chae and Taekyung Kim for helpful discussion.
This work is the result of
research activities (Astrophysical Research Center for the Structure
and Evolution of the Cosmos (ARCSEC)) (A.I.) and
was supported by the Korea Research Foundation Grant funded by
the Korean Government (MOEHRD, Basic Research Promotion Fund)
(KRF-2006-311-C00022) (Y.K.).

\end{document}